\newcounter{myctr}
\def\myitem{\refstepcounter{myctr}\bibfont\noindent\ifnum\themyctr>9\else\phantom{0}\fi\hangindent17pt\themyctr.\enskip}

\documentclass{ws-ijqi}

\begin{document}



\title{Evaluating measures of nonclassical correlation in a multipartite quantum system}

\author{Akira SaiToh${}^1$, Robabeh Rahimi${}^2$, and Mikio Nakahara${}^{3}$}

\address{${}^1$
Department of Systems Innovation, Graduate School of Engineering
Science, Osaka University, Toyonaka, Osaka 560-8531, Japan\\
saitoh@qc.ee.es.osaka-u.ac.jp\\
${}^2$ Research Center for Quantum Computing, Interdisciplinary
Graduate School of Science and Engineering, Kinki University, Higashi-Osaka,
Osaka 577-8502, Japan\\
rahimi@alice.math.kindai.ac.jp\\
${}^3$ Department of Physics, Kinki University, Higashi-Osaka,
Osaka 577-8502, Japan\\
nakahara@math.kindai.ac.jp}

\maketitle


\begin{abstract}
We introduce and compare several measures of nonclassical correlation
defined on the basis of a widely-recognized paradigm claiming that a
multipartite system represented by a density matrix having no product 
eigenbasis possesses nonclassical correlation.
\end{abstract}


\section{Introduction}
Classical/nonclassical separation is a controversial subject.
There have been debates on the definition of nonclassical
correlation of a multipartite system. It is well-known that the
separability paradigm \cite{W89,P96-1,P96-2}
is widely-accepted to define entanglement as a quantum correlation
that cannot be generated from scratch using only local operations and
classical communication (LOCC) (See, e.g., Ref.\ \refcite{PV06}).
There are different paradigms to define nonclassical correlation of a
multipartite system from operational viewpoints.
Bennett {\em et al.} \cite{B99} discussed a certain nonlocality about
locally nonmeasurable separable states.
Ollivier and Zurek \cite{Z02} introduced a measure called quantum
discord defined as a discrepancy of two expressions of a mutual
information that should be equivalent to each other in a classical
information theory. A simple classical/nonclassical separation was given
by Oppenheim {\em et al.} \cite{OH02,H05} which is also widely
recognized. They defined the class of (properly) classically correlated
states that are the states with a product eigenbasis. Its complement is
the class of nonclassically correlated states that are the states
without product eigenbasis. This definition is in accord with their
measure called quantum deficit defined as a discrepancy between the
information that can be localized by applying closed LOCC (CLOCC)
operations and the total information of the system.
The CLOCC protocol allows only local unitary
operations, attaching ancillas in separable pure states, and operations
to send subsystems through a complete dephasing channel. Thus
classically correlated states have vanishing quantum deficit.
Other measures~\cite{G07,S07} were later proposed on the basis of the same
definition of classical/nonclassical correlation.

We aim to evaluate measures based on the following separation of
classical/nonclassical correlations.
\begin{definition}[Oppenheim-Horodecki]
A quantum bipartite system consisting of subsystems A and B
is (properly) classically correlated if and only if it is described
by a density matrix having a biproduct eigenbasis.
\end{definition}
A straightforward extension gives the definition:
\begin{definition}
A quantum multipartite system consisting of subsystems $1,...,m$
is nonclassically correlated if and only if it is described by a density
matrix having no $m$-product eigenbasis.
\end{definition}
The set of classically correlated states is a nonconvex subset of the
set of separable states. Thus a convexity cannot be a property of a
measure of nonclassical correlation.
The natural statement is that a measure $M$ of
nonclassical correlation should satisfy the following conditions:\\
(i) $M=0$ if a system is described by a density matrix having a product
eigenbasis.\\
(ii) $M$ is invariant under local unitary operations.\\
These conditions are considered to be prerequisite hereafter.

In addition, the additivity property should be satisfied if one needs to
compare systems with different dimensions.
There are two ways to define additivity for a measure.
One is defined as follows, which is valid for any measure of bipartite
correlation.
\begin{definition}\label{defadd1}
Let $F(\rho^{\rm AB})_{\rm A|B}$ be a measure of correlation between
subsystems A and B of a bipartite system AB, where ${\rm A|B}$ denotes
splitting between A and B. Then, $F(\rho^{\rm AB})_{\rm A|B}$ is called
an additive measure if and only if the equality
$F(\rho^{\rm AB}\otimes\sigma^{\rm CD})_{\rm AC|BD}
=F(\rho^{\rm AB})_{\rm A|B} + F(\rho^{\rm CD})_{\rm C|D}
$ holds.
\end{definition}
Another definition of additivity may be introduced
for a multipartite measure when we focus on scaling.
\begin{definition}\label{defadd2}
Let us denote a measure of $m$-partite nonclassical correlation by
$M_m(\sigma)$ where $\sigma$ is a density matrix of an $m$-partite
quantum system. First, the measure is fully additive if and only if 
$
M_{m_1\times m_2}(\sigma_1\otimes\sigma_2)=
M_{m_1}(\sigma_1)+M_{m_2}(\sigma_2)
$
with $\sigma_1$ the density matrix of an $m_1$-partite system and
$\sigma_2$ the density matrix of an $m_2$-partite system.
Second, the measure possesses weak additivity if and only if
$
M_{m^n}(\sigma^{\otimes n})=nM_{m}(\sigma).
$
Third, the measure possesses subadditivity if and only if
$
M_{m_1\times m_2}(\sigma_1\otimes\sigma_2)\le
M_{m_1}(\sigma_1)+M_{m_2}(\sigma_2).
$
\end{definition}

In this contribution, we first compare the definitions and properties
of measures of nonclassical correlation in Sec.\ \ref{sec2}. Then we
compare these measures numerically in simple examples in Sec.\
\ref{sec3}. The paper concludes with several remarks.

\section{Nonclassical Correlation Measures}\label{sec2}
Four measures of nonclassical correlation are evaluated for comparison
together with the negativity (an entanglement measure) \cite{neg}.

The first one is an uncertainty remaining for a third party about
the quantum state shared by multiple persons after the third party
receives many reports from them \cite{S07}.
These reports are measurement results on the shared state using local
observables.
Let us consider a density matrix
$\rho^{[1,\ldots,m]}$ of an $m$-partite system.
Consider local complete orthonormal bases $\{|c_j^{[1]}\rangle\}_j, \ldots,
\{|c_x^{[m]}\rangle\}_x$. Suppose the $l$th person locally makes measurements
using the observable $\sum_s s|c_s^{[l]}\rangle\langle c_s^{[l]}|$ and
sends reports to the third party.
Then, a measure of nonclassical correlation is given by
\begin{equation}\label{eqDM}
D(\rho^{[1,\ldots,m]})=
\underset{\mathrm{local~bases}}{\mathrm{min}}
\left(-\sum_{j,\ldots, x}p_{j,\ldots, x}\log_2 p_{j,\ldots, x}\right) -
S_\mathrm{vN}(\rho^{[1,\ldots,m]})
\end{equation}
with $
p_{j,\ldots,x}=\langle c_j^{[1]}|\langle c_k^{[2]}|\cdots \langle c_x^{[m]}|
\rho^{[1,\ldots,m]}|c_j^{[1]}\rangle |c_k^{[2]}\rangle\cdots
|c_x^{[m]}\rangle
$
($S_\mathrm{vN}$ is the von Neumann entropy).
The value of $D(\rho^{[1,\ldots,m]})$ vanishes if $\rho^{[1,\ldots,m]}$
has a (fully) product eigenbasis. In addition,
$D(\rho^{[1,\ldots,m]})$ is invariant under local unitary operations as
is clear from its definition. Furthermore, it is fully additive in terms
of Definition\ \ref{defadd2}. 

The second one is derived from a game of multiple persons sharing a
quantum state. Let us consider an artificial game to find out
eigenvalues of the reduced density matrix of a subsystem from eigenvalues
of the density matrix of the total system.
Suppose that Kate has the $k$th component of an $m$-partite
quantum system.  Let the dimension of the Hilbert space of the $k$th
component be $d^{[k]}$ and that of the Hilbert space of the total system
be $d_{\mathrm{tot}}$. She wants to know the eigenvalues
$\{e_j^{[k]}\}_{j=1}^{d^{[k]}}$ of the reduced density matrix of the
$k$th component. Kate receives $d_{\mathrm{tot}}$ eigenvalues from
Tony who knows all the eigenvalues of the total system. Kate partitions
them into $d^{[k]}$ sets. Summing up elements in individual sets,
she has $d^{[k]}$ mimic eigenvalues $\{\tilde e_j^{[k]}\}$. Thus a
measure of nonclassical correlation for Kate can be
\[
 F_k(\rho^{[1,\ldots,m]})=\underset{{\mathrm{partitionings}}}{{\mathrm{min}}}
\left|\sum_{j=1}^{d^{[k]}}(\tilde e_j^{[k]}\log_2 \tilde e_j^{[k]}
- e_j^{[k]}\log_2 e_j^{[k]})\right|.
\]
We may take the maximum over $k$ to have the measure
\begin{equation*}
 G(\rho^{[1,\ldots,m]})=\underset{{k}}{{\mathrm{max}}}~F_k(\rho^{[1,\ldots,m]}).
\end{equation*}
This is equal to zero if $\rho^{[1,\ldots,m]}$ has a (fully) product
eigenbasis. In addition, it is invariant under local unitary operations
as is clear from its definition.
It is subadditive in terms of Definition\ \ref{defadd2}. 

The third one is a sort of measures introduced by Groisman {\em et
al}.\cite{G07}. This is 
defined in the following way. Consider a bipartite system (AB)
represented by a density matrix $\rho^{\rm AB}$. Then,
(i) Find a basis that diagonalizes the state
${\rm Tr_B}\rho^{\rm AB}\otimes{\rm Tr_A}\rho^{\rm AB}$.~~
(ii) Write $\rho^{\rm AB}$ under the basis found by (i) and delete all
 off-diagonal elements. Denote this state as $\rho'$.~~
(iii) The measure is calculated by a certain distance between
$\rho^{\rm AB}$ and $\rho'$.~~~~
We can take the difference in von Neumann entropy as the distance
function to define the measure:
\[
 D_{\rm G}(\rho^{\rm AB})=S_{\rm vN}(\rho')-S_{\rm vN}(\rho^{\rm AB}).
\]
This measure can be seen as a variant of measure $D$ by taking a fixed
set of local bases instead of searching for the minimum in (\ref{eqDM}).
An extension to the multipartite case is obvious.
This measure satisfies additivity in terms of Definition\ \ref{defadd1}.

To define the fourth one, we can get a clue to construct a measure of
nonclassical correlation from a conventional entanglement measure in
the following way. 
A well known measure of entanglement is negativity \cite{neg},
$N(\rho^{\rm AB})$,
defined as the absolute value of the sum of negative eigenvalues of
$(I^A\otimes\Lambda_{\rm T}^{\rm B})\rho^{\rm AB}$ where $\Lambda_{\rm T}$
is the transposition. This can be in fact regarded as a measure of
nonclassical correlation. Nevertheless, this obviously does not quantify
nonclassical correlation of systems described by separable density
matrices. Instead of negativity, one can define another measure using
the partial transposition:
\[
 K(\rho^{\rm AB})=\sum_x |e_x-\widetilde{e_x}|,
\]
where $e_x$ are the eigenvalues of $\rho^{\rm AB}$ and $\widetilde{e_x}$
are the eigenvalues of $(I^A\otimes\Lambda_{\rm T}^{\rm B})\rho^{\rm
AB}$; both $e_x$'s and $\widetilde{e_x}$'s are aligned in the descending
(or ascending) order. This measure utilizes the fact that $\Lambda_{\rm
T}$ is eigenvalue-preserving while $I\otimes\Lambda_{\rm T}$ is, in
general, not. The partial transposition $I\otimes\Lambda_{\rm T}$
preserves eigenvalues when acting on a density matrix having a biproduct
eigenbasis. In this sense, any
eigenvalue-preserving-but-not-completely-eigenvalue-preserving map
(EnCE, this might be reminiscent of PnCP) can
be used to define a measure of nonclassical correlation. This will be
investigated in detail in our forthcoming contribution \cite{S08}.
One drawback is that a measure in the form of $K$ does not possess
an additivity property, similarly to the case of negativity.
The extension of this measure to a multipartite case can be accomplished
by taking the minimum over all bipartite splittings.

\section{Examples}\label{sec3}
\begin{figure}[pbt]
\begin{minipage}{0.48\textwidth}
\begin{center}
 \scalebox{0.5}{\includegraphics{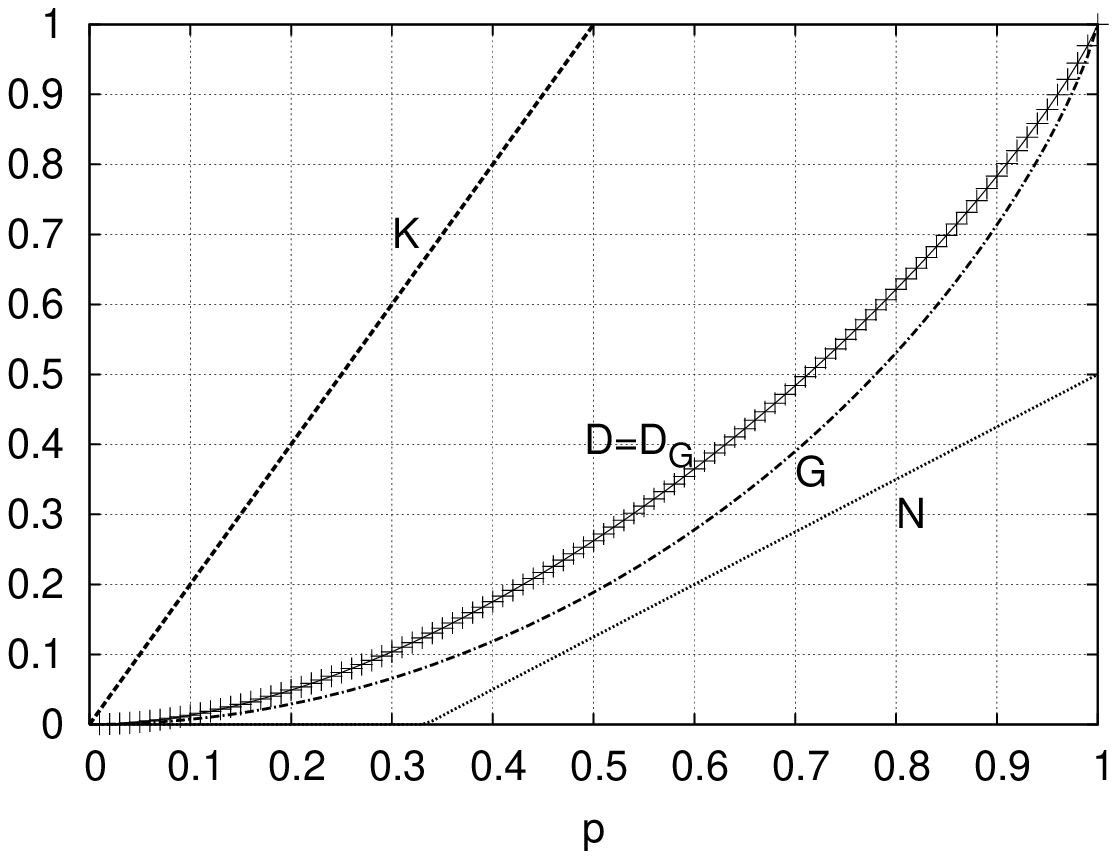}}\\
{\vspace{-2mm}\small$~~~~~~~~~~$(a)}
\end{center}
\end{minipage}
\begin{minipage}{0.48\textwidth}
\begin{center}
 \scalebox{0.5}{\includegraphics{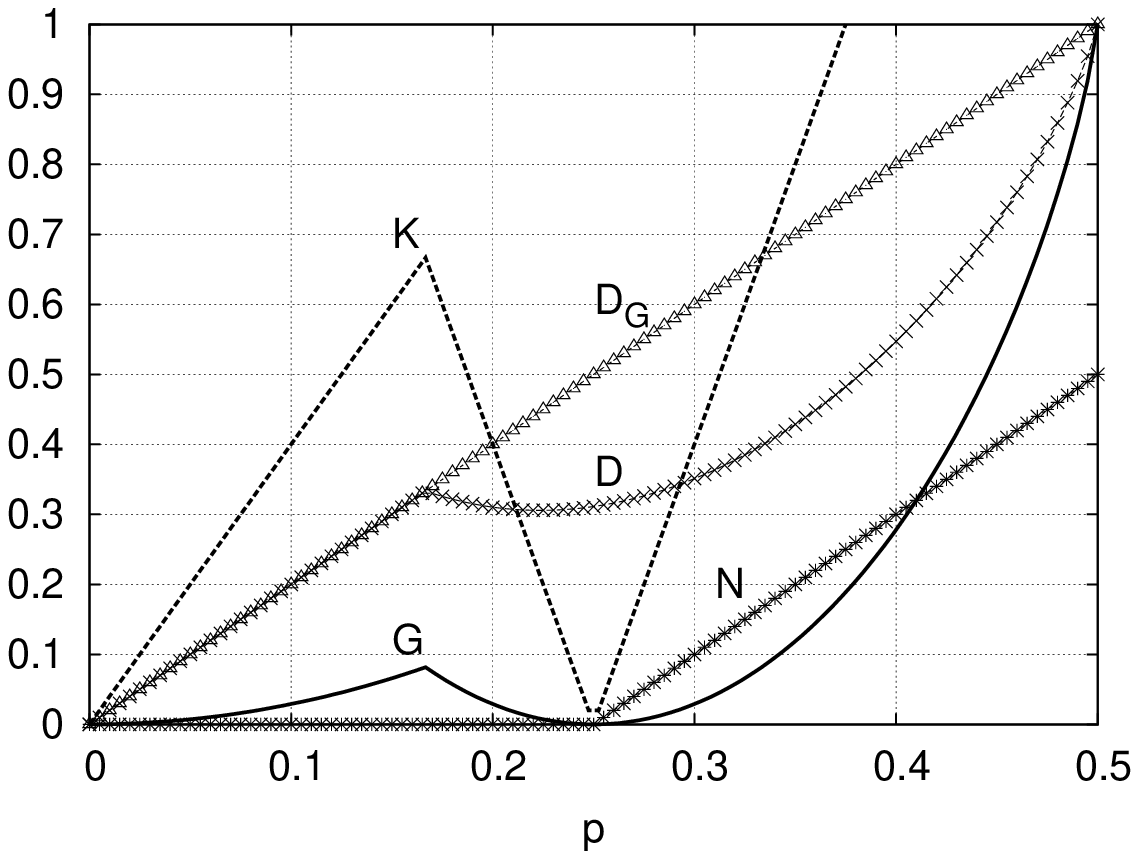}}\\
{\vspace{-2mm}\small$~~~~~~~~~~$(b)}
\end{center}
\end{minipage}\\
\begin{minipage}{0.48\textwidth}
\begin{center}
  \scalebox{0.5}{\includegraphics{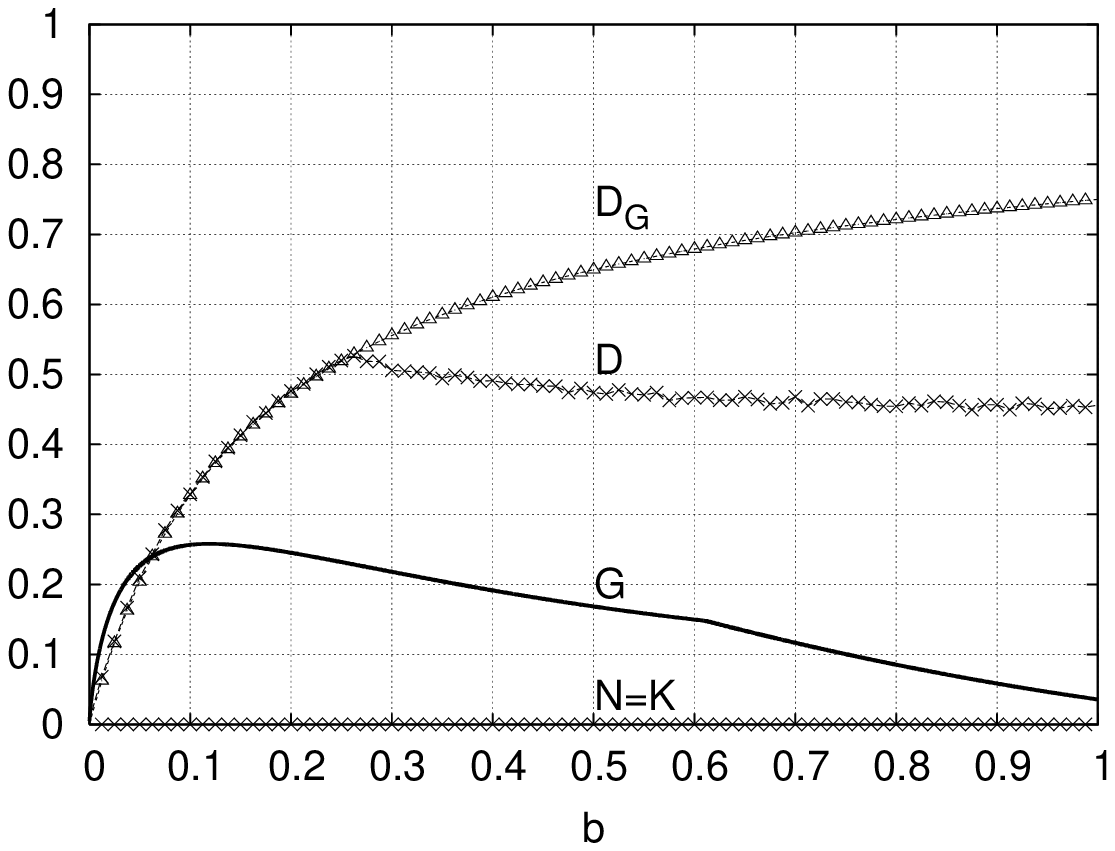}}\\
{\vspace{-2mm}\small$~~~~~~~~~~$(c)}
\end{center}
\end{minipage}
\begin{minipage}{0.51\textwidth}
\begin{center}
\caption{\label{fig}Plots of the measures of nonclassical correlation
introduced in the text against parameters. The target bipartite systems
are those represented by density matrices (a) $\rho_{\rm ps}$,
(b) $\sigma$, and (c) $\sigma_b$.}
\end{center}
\end{minipage}
\end{figure}

We compare the four measures we have seen above together with negativity
in three simple examples of bipartite splitting cases.
The measure $D$ requires a random search of local bases to estimate
its value for a given state. We try $4.0\times 10^4$ randomly generated
bases for each data point in Figs.\ \ref{fig}~(b)~and~(c).
Other measures can be calculated without numerical estimation.

The first example is the pseudo-entangled state of two qubits,
$\rho_{\rm ps}=p|\psi\rangle\langle\psi|+(1-p)I/4$ with
$|\psi\rangle=(|00\rangle+|11\rangle)/\sqrt{2}$.
For this state, we have
$D(\rho_{\rm ps})=D_{\rm G}(\rho_{\rm ps})=
2s[(1+p)/4]-s[(1-p)/4]-s[(1+3p)/4]$ where $s(x)=-x\log_2 x$ ($0\le x\le 1$).
In addition, we have $G(\rho_{\rm ps})=1-H[(1+p)/2]$ where
$H(x)=s(x)+s(1-x)$ is the binary entropy function.
It is also easy to obtain $N(\rho_{\rm ps})=|{\rm min}[0, (1-3p)/4]|$
and $K(\rho_{\rm ps})=2p$.
These results are plotted against $p$ in Fig.\ \ref{fig}~(a).

The second example is the two-qubit density matrix
$\sigma=(1/2-p)(|00\rangle\langle00|+|11\rangle\langle11|)
+2p|\phi\rangle\langle\phi|$ with
$|\phi\rangle=(|01\rangle+|10\rangle)/\sqrt{2}$ and $0\le p\le 1/2$.
This is inseparable for $p>1/4$ since
$(I\otimes\Lambda_{\rm T})\sigma$ has the eigenvalues $1/2-2p$, $p$
(with the multiplicity of two), and $1/2$.
We need to estimate $D(\sigma)$ using a numerical search. As for other
measures, we obtain
$G(\sigma)={\rm min}\{1-H(p+1/2), 1-H(2p)\}$,
$D_{\rm G}(\sigma)=2s(p)-s(2p)$,
$N(\sigma)=|{\rm min}[0, 1/2-2p]|$, and
$K(\sigma)= 4p$ ($0\le p \le 1/6$), $2-8p$ ($1/6 < p \le 1/4$), $8p-2$
($1/4< p \le 1/2$).
These functions are illustrated in Fig.\ \ref{fig}~(b) as functions
of $p$.

The third example is the $8\times 8$ density matrix $\sigma_b$
of the bipartite system $AB$ with the dimensions of the Hilbert spaces,
two for $A$ and four for $B$.
\[\begin{split}
 \sigma_b=&\frac{1}{7b+1}\biggl[
{\rm diag}(b,b,b,b,\frac{1+b}{2},b,b,\frac{1+b}{2})
+b\times(~|0\rangle\langle5|+|1\rangle\langle6|+|2\rangle\langle7|\\
&+|5\rangle\langle0|+|6\rangle\langle1|+|7\rangle\langle2|~)
+\frac{\sqrt{1-b^2}}{2}\times(~|4\rangle\langle7|+|7\rangle\langle4|~)
\biggr]
\end{split}\]
with $0\le b\le 1$.
This was originally introduced by Horodecki \cite{H97} as an
entangled state with positive partial transpose.
In fact $\mathcal{N}(\sigma_b)=0$ although it is inseparable
for $0<b<1$. It is notable that
$I^{\rm A}\otimes\Lambda_{\rm T}^{\rm B}$ does not change the
eigenvalues of $\sigma_b$ and hence $N(\sigma_b)=K(\sigma_b)=0$.
The values of other measures are numerically estimated or
analytically calculated in a straightforward manner while these
are too lengthy to include in the text.
Plots of the measures against $b$ are shown in Fig.\ \ref{fig}~(c).
\section{Concluding Remarks}
There are many different ways to define a measure of nonclassical
correlation. We have seen four of them. The measures $D$ and $D_{\rm G}$
look stable and faithful against changes of parameters among tested
measures as far as we could see in the three simple examples. Further
investigation is required to find desirable properties in addition
to additivity properties. Computational cost should be another factor
to choose a measure. A seemingly natural measure $D$ cannot be used
for a system with a large dimension due to the cost of searching over
all possible local bases. Thus the measures other than $D$ are good
candidates in this sense.

We have found that a more general framework to detect and
quantify nonclassical correlation can be constructed with EnCE,
in analogy with the PnCP map theory, in the process of defining the measure
$K$. Consider a map $\Lambda$ such that $\Lambda$ preserves the
eigenvalues of a density matrix while $I\otimes\Lambda$ in general does
not. It is obvious that $I\otimes\Lambda$ preserves eigenvalues of a
density matrix if it has a biproduct eigenbasis. Thus such $\Lambda$
can be used to detect and quantify nonclassical correlation.
Further investigation in this approach will be
reported elsewhere \cite{S08}.

\section*{Acknowledgments}
AS and RR are independently supported by the Grant-in-Aids for JSPS
Fellows (No. 1808962 and No. 1907329).
MN is supported by ``Open Research Center'' Project
for Private Universities: matching fund subsidy from MEXT.
He is also supported by the Grant-in-Aid
for Scientific Research from JSPS (Grant No. 19540422).

\end{document}